# Personalized rTMS for Depression: A Review

Juha Gogulski[1,3], Jessica Ross[2,1], Austin Talbot[1,2], Christopher Cline[1,2], Francesco L Donati[4], Saachi Munot[1,2], Naryeong Kim[1,2], Ciara Gibbs[5], Nikita Bastin[6], Jessica Yang[1,2], Chris Minasi[1,2], Manjima Sarkar[1,2], Jade Truong[1,2], Corey J Keller[1,2]

[1]Department of Psychiatry & Behavioral Sciences,
Stanford University Medical Center, Stanford, CA, 94305, USA

[2]Veterans Affairs Palo Alto Healthcare System, and the Sierra Pacific Mental Illness, Research, Education, and Clinical Center (MIRECC), Palo Alto, CA, 94394, USA

[3]Department of Clinical Neurophysiology, HUS Diagnostic Center, Clinical Neurosciences, Helsinki University Hospital and University of Helsinki, Helsinki, FI-00029 HUS, Finland

[4]Department of Health Sciences, University of Milan, San Paolo Hospital, Milan, 20142, Italy

[5]Department of Bioengineering, Imperial College London, London, SW7 2AZ, United Kingdom

[6]Departments of Radiology and Orthopedics, Perelman School of Medicine, University of Pennsylvania, 3400 Spruce St, Philadelphia, PA 19104

To whom correspondence should be addressed: Corey Keller
Phone: +1-8025786292
Email: ckeller1@stanford.edu

## Abstract

Personalized treatments are gaining momentum across all fields of medicine. Precision medicine can be applied to neuromodulatory techniques, where focused brain stimulation treatments such as repetitive transcranial magnetic stimulation (rTMS) are used to modulate brain circuits and alleviate clinical symptoms. rTMS is well-tolerated and clinically effective for treatment-resistant depression (TRD) and other neuropsychiatric disorders. However, despite its wide stimulation parameter space (location, angle, pattern, frequency, and intensity can be adjusted), rTMS is currently applied in a one-size-fits-all manner, potentially contributing to its suboptimal clinical response (~50%). In this review, we examine components of rTMS that can be optimized to account for inter-individual variability in neural function and anatomy. We discuss current treatment options for TRD, the neural mechanisms thought to underlie treatment, differences in FDA-cleared devices, targeting strategies, stimulation parameter selection, and adaptive closed-loop rTMS to improve treatment outcomes. We suggest that better understanding of the wide and modifiable parameter space of rTMS will greatly improve clinical outcome.

## 1. Introduction

In modern psychiatry, dysfunctional brain networks are considered the neural substrates of mental health disorders [1]. These networks can be precisely targeted with repetitive transcranial magnetic stimulation (rTMS, referred to here as TMS), a non-invasive treatment with minimal side effects [2, 3]. TMS uses strong (1-2 T) focal magnetic fields to induce electrical currents in neural tissue and elicit neuronal firing [4]. TMS can be applied in single pulses to *probe* cortical excitability or repetitively to *induce* changes in cortical excitability. TMS is currently FDA-cleared for the treatment of obsessive-compulsive disorder [5], smoking cessation [6], migraine [7] and major depressive disorder (MDD, referred to here as depression) [8], with research underway for post-traumatic stress disorder [9, 10], substance use disorders [11], and other neuropsychiatric disorders [12]. For depression, the efficacy and tolerability of TMS have been extensively studied [13–16], but the rate of patients classified as responders or remitters after TMS treatment remains <50% and <20%, respectively, in randomized trials [13]. However, because conventional TMS does not account for significant inter-patient variability that exists, there is likely room for improvement [17]. TMS is currently applied in a one-size-fits-all manner, with minimal individual optimization of stimulation parameters. Furthermore, TMS is currently applied in open-loop fashion, with little-to-no measurements or adjustments during treatment. With this current approach to TMS, we lack the ability to 1) *stratify* patients by predicting responders from non-responders or 2) *optimize* TMS by personalizing treatment parameters.

A wide variety of stimulation parameters can be adjusted within the TMS treatment session, including brain target, pattern, frequency, intensity and number of pulses (Fig 1). Despite this array of options, only a handful of TMS parameters are commonly employed for the treatment of depression in clinical practice: 10 Hz and theta burst stimulation (TBS) to the left dorsolateral prefrontal cortex (dlPFC) and 1 Hz to the right dlPFC. The only TMS parameter that is individualized for depression treatment is intensity, adjusted based on one's motor threshold. This limited utilization of the parameter space is a missed opportunity for precision medicine, and a clear understanding of how each stimulation parameter affects clinical outcome would propel the TMS field forward. Several reasons underlie this gap in knowledge, including 1) the significant time investment required to empirically study each stimulation parameter combination in a randomized clinical trial; 2) the difficulty in measuring neural changes after a single or a few TMS sessions [18]; 3) the lack of understanding between stimulation-induced neural changes on short compared to longer timescales; and 4) the unclear link between these stimulation-induced neural changes and clinical outcome. To move toward a personalized approach with a wider array of treatment patterns, we must 1) develop methodology to measure and monitor neural changes in real-time, 2) study how each stimulation parameter induces acute changes in neural networks, 3) link these acute changes to long-term clinical outcomes, and 4) utilize this knowledge to develop adaptive TMS treatments to drive desired neural changes on a patient-specific level.

In this review, with a focus on enhancing clinical outcome, we examine the TMS parameters that can be individually adjusted during treatment to account for inter-individual variability in brain structure and function. The focus is on the treatment of depression, but similar approaches can be utilized for other neuropsychiatric indications. First, we review current treatment options (including TMS) that are available for patients with treatment-resistant depression (TRD), who do not respond to traditional pharmacological treatments. We next introduce the neural mechanisms and possible sources of interindividual variability in response to TMS. We discuss differences in TMS devices and coils that could be used to personalize TMS treatment [19]. We introduce within- and between-session stimulation parameters that are relevant for optimizing TMS treatment (Fig 1). We focus on human neurophysiology, neuroimaging, and clinical trials pertaining to these factors and provide recommendations for how treatment could be modified to maximize individual efficacy. Finally, we discuss conventional open-loop and novel adaptive, closed-loop TMS and provide suggestions to guide future scientific endeavors towards a treatment paradigm focused on objective biomarkers, real-time monitoring, and adaptive treatment.

## 2. Treatment options for TRD

*2.1. Electroconvulsive therapy (ECT)*

Electroconvulsive therapy (ECT) is considered the gold standard for treating treatment-resistant depression (TRD), providing remission for 50-90% of patients [20]. However, induction of a generalized seizure [21], required anesthesia, and significant side effects including memory loss and confusion [22] preclude ECT as a first-line treatment option.

*2.2. Esketamine*

Esketamine administered as an adjunctive treatment with an oral antidepressant for TRD was approved by the FDA in 2019 [23]. One dose has been shown to relieve depressive symptoms within four hours [24, 25]. Remission rate after nasal administration twice a week for four weeks was observed to be 60% [26]. Common side effects include nausea, anxiety, increased blood pressure, vertigo, dissociation, and hallucinations [27]. Due to the potential for abuse, esketamine can be difficult to obtain and the discontinuation / maintenance process requires further study. In summary, although early esketamine studies demonstrate promising results, the risk of abuse and unclear discontinuation /maintenance process has slowed clinical adoption [28].

*2.3. Experimental Treatment Options for TRD*

Other pharmacologic and non-pharmacologic treatments for TRD include deep brain stimulation (DBS), vagus nerve stimulation (VNS), and psychedelics. DBS uses surgically implanted electrodes to provide focal electrical stimulation to modulate specific brain regions [29]. Two large randomized clinical trials (RCTs) for DBS treatment of TRD failed to show significant differences in response rates between active and control groups [30, 31]. As such, there is little conclusive evidence to date in favor of DBS as a treatment for depression. Vagus nerve stimulation (VNS), FDA-approved for TRD in 2005 [32], uses surgically implanted electrodes to deliver electrical stimulation to the vagus nerve. VNS is thought to treat depression by modulating activity in the default mode network (DMN; [33]). Psilocybin, a serotonergic and glutamatergic modulator thought to be less addictive and neurotoxic than ketamine [34], have shown promising smaller clinical trials [34], with larger studies underway.

*2.4. TMS*

TMS, a non-invasive form of brain stimulation that was FDA-cleared for TRD in 2008, is thought to exert its clinical effects through local and downstream regions connected to the left dlPFC [35, 36]. These neural mechanisms will be reviewed in later sections. TMS has few side effects, including local pain, discomfort, and headache [37]. While a seizure remains a significant potential side effect, current rates of TMS-induced seizures are low (0.31 per 10000 sessions [3, 38]).

*2.5 Summary*

In summary, with an increasing number of interventions available for TRD, more work is needed to determine a standard treatment care pathway. For many of these TRD-targeted interventions, there are concerns about invasiveness, side effect profile, unclear mechanisms of action, limited precision care options, and limited accessibility. TMS has high potential due to its non-invasive nature, minimal side effect profile, numerous personalizable stimulation parameters (Fig 1), and a mechanism of action that is becoming clearer. However, to utilize TMS to its maximal potential and increase its efficacy, it will be necessary to gain significant knowledge around how to optimize and personalize the use of the wide array of TMS parameters.

**3. Neural mechanisms of TMS**

*3.1. Single-pulse TMS*

The principles of TMS are rooted in the theory of electromagnetic induction, whereby a briefly applied, high-intensity pulse of electrical current flowing through the TMS coil generates a strong magnetic field. This rapidly changing magnetic field penetrates the skull and generates an electric field within the brain. Recent computational modeling work suggests that TMS activates axon terminals of all cortical cell types [39]. Stimulation of the commonly studied primary motor cortex [40] results in electrical activations in the associated peripheral muscles (motor evoked potential; MEP). MEPs can be variable within and between individuals [41–43]. This variability is thought to arise from 1) top-down modulation of cortico-spinal excitability [44, 45], 2) phase dynamics of oscillatory brain state [46], 3) random noise at the membrane of the stimulated neurons [47], and 4) slight shifts in TMS coil positioning during recording.

*3.2. Repetitive TMS*

When applied repetitively, rTMS can induce neuronal plasticity by strengthening or weakening cortical excitability in a manner that outlasts stimulation itself [48]. These effects are dependent on the frequency of applied pulses. High frequency (≥ 5 Hz) TMS is thought to be 'excitatory,' and low frequency (≤ 1 Hz) TMS is thought to be 'inhibitory,' although these mechanisms have not been clearly established in humans ([49]; see Section 7), especially outside of motor cortex. These longer-lasting changes in neuronal firing may reflect sustained changes in synaptic plasticity, namely long-term potentiation (LTP) or long-term depression (LTD) [50, 51]. Animal and *in vitro* studies show that the frequency of the applied electrical stimulation pulses determines whether LTP or LTD is induced [52, 53]. In humans, TMS-induced changes in MEP amplitude are thought to reflect LTP [4, 54] or LTD [55–57]. Similar long term plasticity mechanisms may underlie the clinical effects of TMS in non-motor regions such as the prefrontal cortex (PFC) [58, 59], although this idea has not been well-validated.

On a macroscale level, TMS is thought to alleviate symptoms of TRD by normalizing dysfunctional brain networks implicated in depression, including the salience network [1], fronto-parietal network [36] and DMN [60]. Thus, TMS is recognized as a "network therapy," where stimulation applied to one brain region affects local and downstream brain regions [61]. As such, research has shown that the fMRI-guided treatment site in the dlPFC most negatively correlates with the subgenual cingulate cortex, a region implicated in depression [35, 62], may yield improved clinical outcome [61, 63–69]. In summary, to fully personalize TMS treatment, we must better understand how inter-subject differences in macroscale brain networks and microscale LTP / LTD susceptibility play a role in determining clinical outcome.

## 4. TMS devices and coil types

Multiple FDA-cleared TMS devices are commercially available, but a lack of head-to-head studies coupled with the high cost of each device makes it impractical to personalize treatment based on device type. A wide variety of TMS coils are available, each of which vary in stimulation focality and depth [70], the absolute intensity of the induced electric field, and pulse waveform morphology [71]. Regarding coils, the first in use was the circular coil [40], which produces a non-focal ring-shaped electric field, stimulating brain regions under the coil perimeter. To make stimulation more focal, the figure-of-eight coil was then developed [72]. Though numerous adaptations have been described to adjust density, concavity, and heating limitations compared to the figure-of-eight coil [73], this coil type is still the most popular because of a depth-focality trade-off [70]. Another popular coil design is the Hesed coil (H-coil), designed to stimulate deeper brain structures [74]. While other designs exist to address various problems of focality, depth, and target control [12], including the double cone coil, cap coil, D-shape coil and C-core coil, it is generally accepted that no TMS coil can effectively achieve deep and focal stimulation simultaneously [70, 75]. In summary, the wide variety of coil shapes are intriguing and represent a potential 'parameter' that can be personalized, but situations in which one should use a specific coil type are not yet clear. Though various coil types have been studied with computational models and simulations [76–78], real-world human studies are needed to better understand how different coil types and pulse waveforms affect neurophysiological and clinical outcomes.

## 5. Stimulation location

While the left dlPFC is the main treatment site for TMS, how to most effectively target the optimal dlPFC subregion for each subject remains an area of active research [61, 64, 79]. Scalp-based targeting methods (5, 5.5, and 6 cm and Beam F3) were the earliest and are still the most widely utilized techniques due to their low cost and ease of use. The 5 cm rule describes a scalp location 5 cm anterior to primary motor cortex. As this method does not account for variations in head dimensions and occasionally leads to stimulation of premotor rather than prefrontal cortex [61, 64, 80–82], more anterior and lateral targeting including the 5.5 and 6 cm rule were proposed [61, 64, 79]. An alternative method, termed the Beam F3 method, places the target 1 cm anterolateral to the F3 electroencephalography (EEG) electrode [83]. Although not yet validated by a prospective, double-blind study, recent work suggests that the Beam F3 method improves reproducibility, precision, and inter- and intra-rater reliability of TMS treatments compared to older 5-6 cm methods [61, 79].

Although clinically useful, scalp-based methods are less reliable than using an individual's MRI to target the dlPFC [84, 85], often missing the target by 1-2 cm [85]. In addition to macro-anatomical differences [85], inter-individual differences in functional connectivity may partially contribute to the variability in clinical response across subjects [64, 65, 86]. Thus, one could potentially improve clinical response by incorporating fMRI or EEG, modalities that capture brain network activity, into dlPFC targeting algorithms by accounting for these individual differences. While promising, larger prospective, controlled studies between scalp-, EEG-, MRI-, and fMRI-based targeting are needed to more definitively determine differences in neurophysiological and clinical outcomes between these methods.

In addition to the left dlPFC, alternative treatment targets have been studied. Perhaps the most promising alternative target is the right dlPFC where 1 Hz stimulation is typically applied. 1 Hz right dlPFC TMS delivers similar clinical results as high-frequency TMS to the left dlPFC [2, 15, 81, 87–91]. TMS to the dorsomedial prefrontal cortex (dmPFC), a central node of emotion and behavioral regulation networks [61, 88, 92, 93], is safe, tolerable, and in small studies produces response and remission rates comparable to dlPFC TMS [92]. TMS of the right orbitofrontal cortex (OFC), an area involved in nonreward, also produces antidepressant effects and mood improvements [94, 95]. In summary, alternative targets include right dlPFC, dmPFC, and OFC, but more studies are needed before widespread adoption and consideration of treatment target as a stimulation 'parameter' to personalize.

**6. Stimulation intensity**

*6.1. Determination based on motor threshold*

The most common approach for determining TMS intensity is to assess the subject's resting motor threshold (rMT) [96]. The rMT is defined as the minimum intensity needed to produce an MEP of at least 50 µV in 5 out of 10 single TMS pulses delivered to the primary motor cortex for a hand muscle (abductor pollicis brevis, APB; or first dorsal interosseous, FDI) at rest [97]. The rMT assessment is fast, safe, and usually well tolerated [98]. It indirectly accounts for inter-individual differences in scalp-to-cortex distance and cortical excitability [99]. However, recent evidence has challenged the classical assumption that rMT has high intra-subject stability. Individual rMT is affected by volitional [100] and attentional mechanisms [101] and differs between stimulation devices [102]. Moreover, the rMT may change throughout the course of a standard TMS treatment [103].

Because scalp-to-cortex distance and local dlPFC excitability may differ from the primary motor cortex [104], determination of dlPFC treatment intensity using rMT may result in considerable heterogeneity in effective stimulation intensity across subjects [105]. Using an intensity relative to the rMT may result in under- or over-dosing of TMS [106] and contribute to variable treatment responses [107]. A simple method to partially address this issue is to adjust the stimulation intensity according to the individual scalp-to-dlPFC distance, as measured on the subject's structural MRI scan [108, 109]. Such an adjustment has been suggested to be particularly useful in older patients [110], where cortical atrophy may cause large scalp-to-dlPFC distances [111]. However, this method does not take into account possible differences in cortical excitability between the dlPFC and the primary motor cortex, and as such has limitations as a dosing strategy [107, 112].

*6.2. Determination based on E-field modeling*

Recent advances in neuroimaging and brain modeling have enabled methodology to accurately estimate the electric field (E-field, expressed in V/m) induced on the cortex by TMS [113–116]. Using the subject's individual structural MRI, E-field modeling techniques take into consideration the conductivity of head tissues (scalp, skull, cerebrospinal fluid, etc), scalp-to-cortex distance, and gyral folding pattern, a major source of variability between individuals in the dlPFC [86, 117]. E-field modeling reduces the stimulator-dependent variability in TMS intensity [102]. Furthermore, E-field modeling can account for coil shape and orientation, which greatly affect the induced electric field at the cortex [117, 118]. Lastly, by employing personalized tractography, the induced E-field within specific white-matter tracts can be calculated [119]. If E-field modeling is not used, it can result in underdosing when dlPFC is stimulated [120]. Thus, E-field modeling can provide more precise and controlled dosing for TMS compared to rMT-based methods.

*6.3. Determination using real-time TMS-EEG monitoring*

One way to monitor stimulation of neural tissue in the dlPFC is to combine TMS with simultaneous EEG (TMS-EEG). TMS-EEG has been employed to probe brain responses to TMS in healthy subjects [121] and patients with neuropsychiatric disorders [122–124] including depression [125]. TMS-EEG has also been used to characterize brain changes after TMS treatment [126]. The main functional readout of TMS-EEG, the TMS-evoked potential (TEP), is relatively stable [127] and can be visualized in real time [128]. This allows the TMS operator to modify the stimulation parameters (intensity, coil orientation, etc.) based on a direct functional readout from the stimulated brain area. Future studies should investigate the effects of EEG-informed TMS dosing strategies on treatment outcomes compared to conventionally dosed TMS.

In conclusion, while adjusting the intensity of TMS as a fixed percentage of the individual patient's rMT is safe, fast, and affordable, it might lead to under or overdosing of TMS, potentially increasing the variability of clinical response [106]. Novel strategies employing E-field modeling and EEG represent promising avenues for personalizing TMS treatment based on brain imaging or electrophysiological readout. Moreover, TMS-EEG might be useful for estimating plasticity induced by TMS. Nevertheless, comparisons between rMT-based, E-field-based, and EEG-based dosing methods in clinical populations are needed.

## 7. Stimulation Frequency & Pattern

While TMS is capable of a wide variety of stimulation frequencies and patterns, only a narrow range of TMS parameters have been utilized. Standard TMS protocols can be broadly categorized into high frequency ($\geq$ 5 Hz, 'excitatory') and low frequency ($\leq$ 1 Hz, 'inhibitory') [49]. While multiple frequencies (1, 5, 10, 20 Hz) have been shown to be clinically effective for depression [129–136], to our knowledge only 5 Hz and 10 Hz have been directly compared in a large RCT [137]. In contrast, TBS is a patterned protocol in which high frequency (50 Hz) bursts of three pulses are delivered at 5 times per second [138]. Intermittent TBS (iTBS), in which two-second trains of bursts are repeated every 10 seconds, is thought to be 'excitatory' and induce LTP, whereas continuous TBS (cTBS) is thought to be 'inhibitory' and induce LTD [139]. Perhaps the most commonly used iTBS protocol delivers 600 pulses within a session, lasting 3 minutes [138]. In a seminal and field-changing study, Blumberger et al demonstrated that iTBS to the left dlPFC (600 pulses, 3 minutes) was not inferior to 10 Hz (3000 pulses, 17.5 minutes) for reducing depressive symptoms in TRD [140]. Largely due to this study, TBS was FDA-cleared to treat TRD in 2018. And due to the brevity of TBS, it has become standard in many TMS clinics.

## 8. Accelerated TMS protocols

Although TMS is typically applied once daily, recent work suggests that multiple treatments per day may be as effective and could improve overall access due to a shorter total number of weeks in treatment [141]. To date, the optimal number of treatments per day and the optimal time interval between treatments are unknown. Regarding the number of treatments per day, TMS applied twice [142] and even up to ten times daily [141, 143] has been shown to be clinically effective. These accelerated procedures have used 10 Hz [144–147], 20 Hz [148, 149] and iTBS [143, 150–152] protocols. In another field-changing study, Cole et. al. demonstrated impressive clinical efficacy of ten times daily left dlPFC TBS for one week (50 total treatments) compared to sham [141]. Other accelerated protocols such as with 1 Hz or cTBS appear to be safe but have limited clinical data [153, 154]. Accelerated protocols vary widely with respect to sessions per day (2-10), inter-session interval timing (12-120 minutes), treatment duration (2-10 days), and total number of pulses delivered per day. Outside of accessibility to the patient (which is a significant advantage), it is too early to determine the clinical benefits regarding the effects of the number of treatments per day, inter-treatment interval, and total number of treatments. As such, additional RCTs comparing the clinical efficacy within accelerated protocols and between accelerated and standard protocols are necessary. In summary, accelerated TMS may represent the future of the field – number of treatments per day, inter-treatment intervals, and total number of treatments – can be viewed as additional stimulation 'parameters' to personalize and maximize clinical response.

## 9. Closing the Loop: Towards adaptive, personalized TMS treatment

*9.1. Current state of closed-loop stimulation for TRD*

Intuitively, a fully personalized TMS treatment would provide more effective and faster clinical response. To fully personalize TMS, one needs to 1) identify the individual brain network of interest and its dysfunctionality, 2) determine the metric (biomarker) that tracks brain change and relates to clinical outcome, and 3) adapts TMS in closed-loop fashion to maximize the degree of biomarker change in the direction back to the 'healthy' state. Recent invasive work in a depressed patient undergoing experimental surgery demonstrated proof of concept (N=1) of this approach. Here, the investigators first mapped mood symptoms to amygdala activity. Then, using amygdala activity as a brain biomarker they guided deep brain stimulation in closed-loop fashion to alleviate depression symptoms [155, 156]. For broader access, translation to non-invasive modalities like TMS is needed. Non-invasively, TMS timed to the specific phase of ongoing brain oscillations modulate cortical excitability better than non-phase timed TMS [157–159]. While promising, fully closed-loop TMS treatment, where stimulation parameters are modified based on treatment-related feedback from brain activity, has not been implemented. Several methodological constraints currently limit this type of individualized closed-loop TMS treatment, as outlined below.

*9.2. Challenges of targetable biomarker identification*

Although several studies have reported promising findings [126, 160], we still lack a reliable, reproducible non-invasive brain biomarker for TMS treatment. A targetable biomarker that can be used for closed-loop stimulation must fulfill stringent criteria: 1) the biomarker must be *reproducible* in new individuals. In other words, this biomarker must be indicative of depression in the larger population, not just the individuals in the study. The clinician must be confident that in a new patient the biomarker will properly indicate treatment response. 2) The biomarker must have *clinical significance*, and to do so the relationship between the biomarker and depression must be substantial. Statistical significance traditionally used in science is insufficient because a sufficiently large sample size can reveal the most minute effects. These criteria are difficult to fulfill. Addressing 1) requires that the biomarker is evaluated in *new individuals* not used to teach the biomarker. This is becoming more standard, but is difficult as it requires sample sizes of a relatively large number of individuals. Addressing 2) can be done by evaluating based on p-values, biomarkers should be selected that explain a substantial portion of the variance in the signal. Without a reliable biomarker, one cannot reliably link ongoing brain activity to clinical outcome in real time. To properly implement closed-loop TMS, developing a robust and reproducible depression biomarker is a top priority. Several potential depression biomarkers have been developed [126], but none to date have been validated on external datasets.

*9.3. Closing the loop: Control theory, reinforcement learning, and Bayesian optimization*

Once a predictive brain biomarker has been identified, the question will be how to adapt TMS treatment parameters in real time based on the state of the biomarker. There are three general mathematical frameworks that have been considered for closed-loop stimulation: *control theory* [161], *reinforcement learning* [162], and *Bayesian optimization* [163]. Each of these frameworks arose from different fields to address related but distinct problems.

*Control theory* seeks to maintain a dynamical system (the brain) in a desired state (healthy) in an optimal fashion [164]. For TMS, the algorithm would measure and feed changes in the brain biomarker back to a *controller* that would modify a subsequent *process value* (in this case stimulation parameters such as frequency, intensity, duration). The overarching goal with this algorithm would be to modify the brain biomarker toward the healthy state and thus minimize the number of sessions or total pulses required to achieve clinical remission. Unfortunately, most types of control theory require a precisely defined relationship between the stimulation parameters and biomarker change. While this is possible in areas such as robotics, this requirement makes applying such techniques to psychiatry and neuroscience difficult.

*Reinforcement learning* was developed largely as part of artificial intelligence to teach computers to accomplish tasks in an optimal manner. This requires three components: 1) a set of possible *actions* (stimulation parameters), 2) an *environment* that responds to actions (the brain), and 3) a *reward* function (change in brain biomarker strength). Given these components, reinforcement learning seeks to learn a *policy* (stimulation strategy) to maximize the accumulated *reward* (i.e., maximize brain biomarker change). This learning balances exploitation of the best known parameter value with exploration of potentially better alternatives in a mathematically optimal manner [165]. Reinforcement learning has been successfully used in a variety of applications, such as self-driving cars [166], dynamic health treatments [167], and medication delivery [168].

*Bayesian optimization* was developed in machine learning to optimize parameters used in complicated statistical models by minimizing a loss function [169]. Recently, this method has been successfully utilized to determine the TMS coil orientation that maximizes the amplitude of single TMS-EEG responses in real time [170]. In theory, the same methodology can be used to optimize treatment efficacy of TMS by determining the stimulation parameters that maximize biomarker change in the direction of a healthy individual. Bayesian optimization in TMS starts by creating a *surrogate model* (an approximation to the patient's stimulation response) and evaluates the function at various parameter configurations (i.e. different frequencies or intensities) to increase the accuracy of this approximation (Fig 2). These configurations are chosen based on a second function, known as the *acquisition function*, which balances evaluating new stimulation parameter values (*exploration*) and fine-tuning well-characterized parameter values (*exploitation*) [163]. In contrast to *reinforcement learning* which attempts to maximize the total response (i.e. total change in brain biomarker), *Bayesian optimization* seeks only to discover the *optimal* stimulation procedure by the end of training.

Each of these three methods possess distinct advantages and disadvantages. *Control theory* provides rigorous mathematical guarantees but as it requires a clear understanding between the controller and process values, such as in robotics, it would be limited in the case of closed-loop TMS where little is known between the controller and process value relationship. *Reinforcement learning* can develop a policy in an unknown environment that maximizes the reward over the entire time-period, but developing these complicated policies requires an inordinate amount of data, at times months of training corresponding to millions of observations [171]. *Bayesian optimization*, by contrast, is designed to provide reasonable answers even with limited numbers of samples and as such we recommend Bayesian optimization as a promising method for closed-loop TMS. In summary, as TMS is quickly moving towards precision medicine approaches, the choice and understanding of these algorithms and biomarkers will be critical in the development of adaptive, closed-loop TMS for depression and other neuropsychiatric disorders.

## 10. Discussion

In this review, we examine the TMS parameters that can be personalized to account for inter-individual variability in brain structure and function. Personalized TMS holds great promise to improve TMS treatment for depression, both in terms of clinical efficacy and time burden on patients (number and length of visits). Parameters that could be adjusted for personalization are depicted schematically in Figure 1. These include: 1) within-session parameters (target, frequency, pattern, intensity and number of pulses), 2) between-session parameters (inter-session spacing, number of treatments per day and total number of treatments), and 3) hardware parameters (coil type and pulse waveform). While parameters associated with TMS hardware are difficult to modify in real-time, all within- and between-session TMS parameters are practically modifiable in real-time. However, critical limitations to implement include a lack of 1) methodology to track brain network changes in real-time and 2) understanding of the relationship between stimulation parameters and brain network changes. Several groups are currently investigating these important questions, and proof of concept studies to modify some of these stimulation parameters are promising [156, 170]. We remain optimistic that the next years of research will see enormous gains in understanding the relationship between stimulation parameters, brain biomarkers, and clinical outcome, and as such will lead to the development of personalized algorithms that efficiently and effectively close the loop for fully personalized TMS treatment.

Several aspects of TMS personalization were not discussed here but warrant further review. First, another form of TMS personalization includes stratifying treatment options based on *clinical subtyping* [172] or accounting for inter-individual differences in sets of symptoms within the same clinical diagnosis [17], as well as *biological subtyping* [173] accounting for inter-individual differences in brain networks [174]. Second, a discussion around the optimal *brain state* during TMS is warranted. Should we perform psychotherapy [175] or a cognitive task before, during, and/or after TMS to activate relevant brain networks to enhance the efficacy of TMS? These decisions would likely have large effects on how TMS modulates brain activity and clinical outcome, but to date have not been rigorously studied. Finally, a discussion about the feasibility of implementing sophisticated hardware and software associated with neuronavigation or adaptive TMS in the clinic warrants discussion. Indeed, risk / benefit discussions around time commitment, financial impact and limitations, accessibility, and efficacy are necessary and may vary for each clinic and even each patient. In the end, treatment options that are effective, fast, user-friendly, and inexpensive will be most desirable for clinics and patients.

To help guide further research investigation of the many modifiable stimulation parameters (Fig 1), we provide some general recommendations after reviewing the literature for clinicians interested in more personalized TMS treatment for depression. Consider these points in designing your protocols:

- Figure-of-eight shaped coils provide a good depth / focality tradeoff out of coils currently available.
- The left dlPFC brain target is the most commonly used and studied target for depression. However, future studies on optimal coil location and angle within the left dlPFC as well as promising alternative targets (right dlPFC, dmPFC, and OFC) are needed.
- MRI or fMRI-guided TMS may be more effective than scalp-based navigation but prospective comparison studies examining added time, cost, and clinical efficacy are needed.
- Choosing TMS dosage based on rMT can be improved by applying a scalp-to-cortex adjustment based specifically on the dlPFC target. If available, E-field modeling should be considered, as this technique takes into consideration scalp-to-cortex distance, gyral folding pattern, and coil shape and orientation.
- Once daily 10 Hz TMS is the most common and studied protocol for depression, but TBS is quickly becoming mainstream due to its clinical equipoise with 10 Hz and reduced time requirements.
- Accelerated TMS protocols provide an intriguing set of inter-session parameters to personalize including number of treatments per day and spacing between treatments, but further testing is needed to understand how these parameters affect clinical efficacy.

## **Acknowledgements**

We would like to acknowledge the generous contributions of the members of the Personalized Neurotherapeutics Laboratory (kellerlab.stanford.edu) for helpful feedback on the manuscript.

This research was supported by the National Institute of Mental Health under award number R01MH126639, R01MH129018, and a Burroughs Wellcome Fund Career Award for Medical Scientists (CJK).

JG was supported by personal grants from Orion Research Foundation and the Finnish Medical Foundation. JMR was supported by the Department of Veterans Affairs Office of Academic Affiliations Advanced Fellowship Program in Mental Illness Research and Treatment, the Medical Research Service of the Veterans Affairs Palo Alto Health Care System and the Department of Veterans Affairs Sierra-Pacific Data Science Fellowship.

## **Disclosures**

CJK holds equity in Alto Neuroscience, Inc. All other authors have nothing to disclose.

**Figures and Legends**

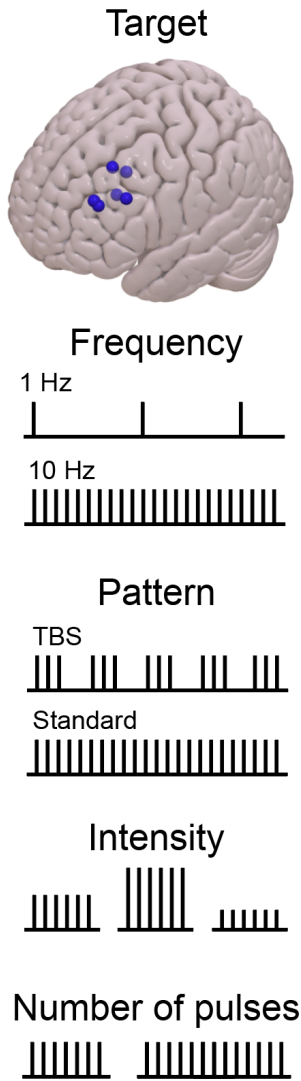
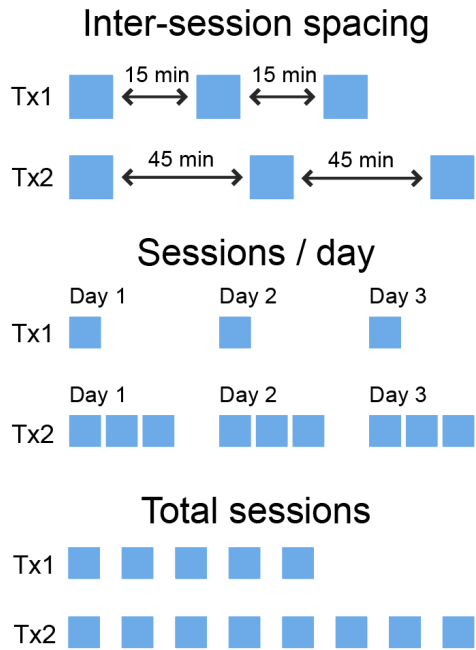
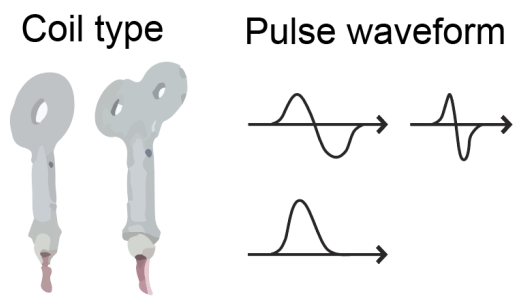

**Figure 1 - TMS parameters**. Within-session and between-session parameters can potentially be personalized and adjusted in real-time using adaptive TMS, whereas hardware parameters are more difficult to modify in real-time.

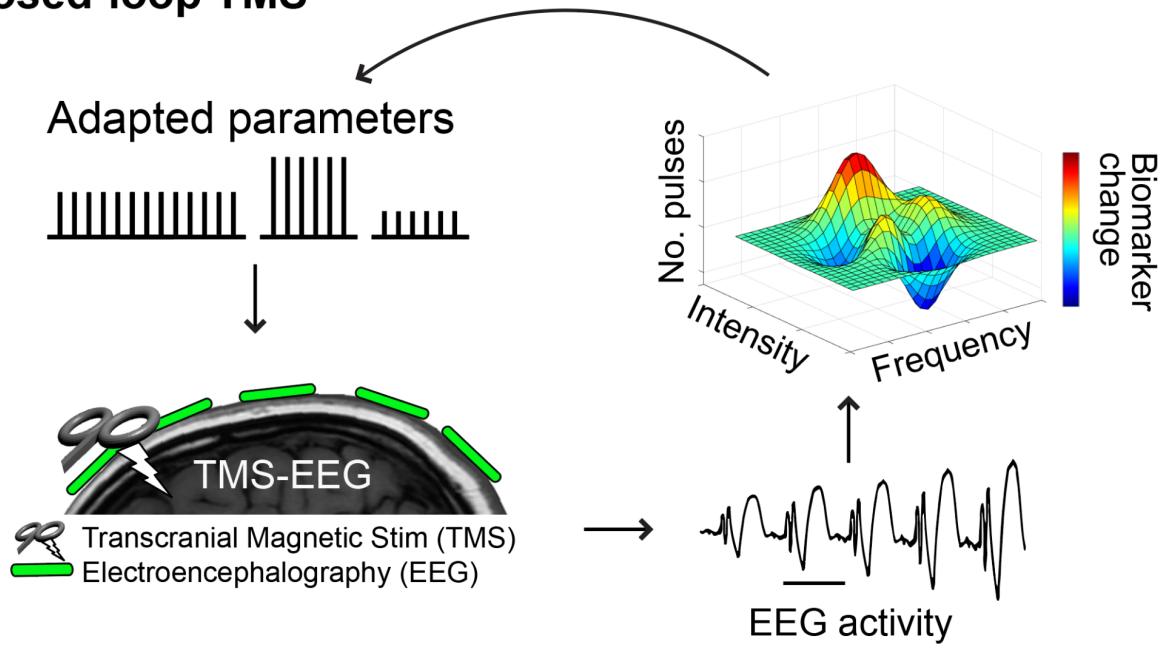

**Figure 2 - Closed-loop TMS.** Closed-loop TMS involves real-time biomarker monitoring and feedback to adapt TMS parameters in order to maximize the desired biomarker change. *Bayesian optimization* could be used to discover the optimal TMS parameter combination with a limited number of trials.